\documentclass[twoside,twocolumn,9pt]{article}
\usepackage{extsizes}
\usepackage[super,sort&compress,comma]{natbib} 
\usepackage[version=3]{mhchem}
\usepackage[left=1.5cm, right=1.5cm, top=1.785cm, bottom=2.0cm]{geometry}
\usepackage{balance}
\usepackage{times,mathptmx}
\usepackage{sectsty}
\usepackage{graphicx} 
\usepackage{lastpage}
\usepackage[format=plain,justification=justified,singlelinecheck=false,font={stretch=1.125,small,sf},labelfont=bf,labelsep=space]{caption}
\usepackage{float}
\usepackage{fancyhdr}
\usepackage{fnpos}
\usepackage[english]{babel}
\addto{\captionsenglish}{%
  
}
\usepackage{array}
\usepackage{droidsans}
\usepackage{charter}
\usepackage[T1]{fontenc}
\usepackage[usenames,dvipsnames]{xcolor}
\usepackage{setspace}
\usepackage[compact]{titlesec}
\usepackage{hyperref}
\usepackage{subcaption}
\usepackage[detect-all]{siunitx}

\usepackage{epstopdf}

\definecolor{cream}{RGB}{222,217,201}

\begin{document}

\pagestyle{fancy}
\thispagestyle{plain}
\fancypagestyle{plain}{

\renewcommand{\headrulewidth}{0pt}
}

\makeFNbottom
\makeatletter
\renewcommand\LARGE{\@setfontsize\LARGE{15pt}{17}}
\renewcommand\Large{\@setfontsize\Large{12pt}{14}}
\renewcommand\large{\@setfontsize\large{10pt}{12}}
\renewcommand\footnotesize{\@setfontsize\footnotesize{7pt}{10}}
\makeatother

\renewcommand{\thefootnote}{\fnsymbol{footnote}}
\renewcommand\footnoterule{\vspace*{1pt}%
\color{cream}\hrule width 3.5in height 0.4pt \color{black}\vspace*{5pt}} 
\setcounter{secnumdepth}{5}

\makeatletter 
\renewcommand\@biblabel[1]{#1}            
\renewcommand\@makefntext[1]%
{\noindent\makebox[0pt][r]{\@thefnmark\,}#1}
\makeatother 
\renewcommand{\figurename}{\small{Fig.}~}
\sectionfont{\sffamily\Large}
\subsectionfont{\normalsize}
\subsubsectionfont{\bf}
\setstretch{1.125} 
\setlength{\skip\footins}{0.8cm}
\setlength{\footnotesep}{0.25cm}
\setlength{\jot}{10pt}
\titlespacing*{\section}{0pt}{4pt}{4pt}
\titlespacing*{\subsection}{0pt}{15pt}{1pt}

\fancyfoot{}
\fancyfoot[RO]{\footnotesize{\sffamily{1--\pageref{LastPage} ~\textbar  \hspace{2pt}\thepage}}}
\fancyfoot[LE]{\footnotesize{\sffamily{\thepage~\textbar\hspace{3.45cm} 1--\pageref{LastPage}}}}
\fancyhead{}
\renewcommand{\headrulewidth}{0pt} 
\renewcommand{\footrulewidth}{0pt}
\setlength{\arrayrulewidth}{1pt}
\setlength{\columnsep}{6.5mm}
\setlength\bibsep{1pt}

\makeatletter 
\newlength{\figrulesep} 
\setlength{\figrulesep}{0.5\textfloatsep} 

\newcommand{\topfigrule}{\vspace*{-1pt}%
\noindent{\color{cream}\rule[-\figrulesep]{\columnwidth}{1.5pt}} }

\newcommand{\botfigrule}{\vspace*{-2pt}%
\noindent{\color{cream}\rule[\figrulesep]{\columnwidth}{1.5pt}} }

\newcommand{\dblfigrule}{\vspace*{-1pt}%
\noindent{\color{cream}\rule[-\figrulesep]{\textwidth}{1.5pt}} }

\makeatother

\twocolumn[
\begin{@twocolumnfalse}
\sffamily

 \noindent\LARGE{\textbf{Bacteria driving droplets$^\dag$}} \\

\noindent\large{Gabriel Ramos,$^{\ast}$ Mar\'ia Luisa Cordero, and Rodrigo Soto} \\

\noindent\normalsize{We confine a dense suspension of motile \textit{Escherichia coli} inside a spherical droplet in a water-in-oil emulsion, creating a "bacterially" propelled droplet. We show that droplets move in a persistent random walk, with a persistence time $\tau\sim\SI{0.3}{\second}$, a long-time diffusion coefficient $D\sim\SI{0.5}{\micro\meter^2/\second}$, and an average instantaneous speed $V\sim \SI{1.5}{\micro\meter/\second}$ when the bacterial suspension is at the maximum studied concentration. Several droplets are analyzed, varying the drop radius and bacterial concentration. We show that the persistence time, diffusion coefficient and average speed increase with the bacterial concentration inside the drop, but are largely independent of the droplet size. By measuring the turbulent-like motion of the bacteria inside the drop, we demonstrate that the mean velocity of the bacteria near the bottom of the drop, which is separated from a glass substrate by a thin lubrication oil film, is antiparallel to the instantaneous velocity of the drop. This suggests that the driving mechanism is a slippery rolling of the drop over the substrate, caused by the collective motion of the bacteria. Our results show that microscopic organisms can transfer useful mechanical energy to their confining environment, opening the way to the assembly of mesoscopic motors composed of microswimmers.} \\


\end{@twocolumnfalse} \vspace{0.6cm}

]

\renewcommand*\rmdefault{bch}\normalfont\upshape
\rmfamily
\section*{}
\vspace{-1cm}


\footnotetext{Departamento de F\'\i sica, FCFM, Universidad de Chile, Av. Blanco Encalada 2008, Santiago, Chile.; E-mail: gabriel.p.ramos.p@gmail.com}

\footnotetext{\dag~Electronic Supplementary Information (ESI) available: [details of any supplementary information available should be included here]. See DOI: 00.0000/00000000.}




Active systems, such as animal flocks, motile bacteria, active colloids, or biofilaments with associated molecular motors, are characterized by the injection of mechanical energy at the particle level~\cite{Marchetti2013}. At microscopic scales, these systems are highly affected by thermal and chemical noise and by the random orientations and positions of the active constituents. Despite this inherent randomness, cells are able to extract useful and directed work from molecular motors to drive internal processes, such as mitosis, cellular motility, or rotation of flagella~\cite{Bustamante2001}. Somewhat inspired by these examples, extraction of useful work in various systems of active matter has interested several researchers recently. For example, microtubules and kinesin motors encapsulated in a microscopic drop were used to move and deform the confining drop~\cite{Sanchez2012, Keber2014}, and suspensions of motile bacteria have been used to rotate microscopic gears~\cite{Sokolov2010, DiLeonardo2010}. Bacteria tend to align in a shear flow and in this situation they have been shown to exert work on the suspending fluid and decrease its viscosity~\cite{Lopez2015}.

Take the example of bacterial baths. Modeled as self-propelled particles, individual bacteria can be characterized by a thrust force of magnitude $f \sim \eta \ell v_0$, where $v_0$ and $\ell$ are the characteristic speed and size of the bacterium, and $\eta$ the viscosity of the ambient fluid. Hence, in a suspension with a volumetric concentration $n$ of bacteria, there exists a potential to extract a mechanical power $P \sim n f v_0$ per unit volume. Using typical values for {\it Escherichia coli} ({\it E. coli}), $f \approx \SI{0.5}{\pico\newton}$~\cite{Drescher2011} and $v_0 \approx \SI{20}{\micro\meter/\second}$ in aqueous media, and for a relatively dense suspension of $n \approx 10^{10}$~bact/\si{\milli\liter} ($\sim 1\%$ in volume fraction), one obtains $P \sim \SI{0.1}{\micro\watt/\milli\liter}$.

In this work we extract part of this work to move a sub-millimetric emulsion drop filled with a suspension of motile {\it E. coli}. Bacteria are confined within the drop, thus forming a ``motor made of motors''~\cite{Sanchez2012}. We show that bacterial flows are able to transfer movement to the drop, which performs a persistent random walk. Previous works on bacterial baths confined within emulsion drops have shown organized collective motion of bacteria~\cite{Wioland2013} and their accumulation at the drop interfase~\cite{Vladescu2014}. Both ingredients are key in our work. At high concentration, bacteria organize in turbulent-like, chaotic collective motions known as active turbulence~\cite{wensink2012meso, Gachelin2014}. Also, bacteria confined inside the drops are unable to escape or deform the drop surface, as can be easily understood from an energy balance. The energy required to create a bump of size comparable to the bacterial body in the drop surface is $\sim \gamma/\ell^2$, with $\gamma$ the interfacial tension, while the energy that a bacterium spends by swimming the same distance is $f \ell \sim \eta \ell^2 v_0$. The ratio between these two energies is of the order of the capillary number, $Ca = \eta v_0/\gamma \sim 10^{-5}$, considering a typical water/oil interfacial tension in the presence of surfactants, $\gamma \sim \SI{1}{\milli\newton/\meter}$. As a result, bacteria swimming near a typical water/oil interface feel a rigid boundary and thus behave similarly than when they swim near a solid wall~\cite{Lauga2006} or a free surface~\cite{DiLeonardo2011}; they interact with their hydrodynamic image and accumulate at the interface. This accumulation near the drop interfase~\cite{Berke2008, Vladescu2014} can enhance the interaction of the bacterial flows in the drop and the fluid surrounding the drop.

Microswimmers such as {\it E. coli} swim by exerting a force dipole on the ambient fluid, and are thus force and torque free~\cite{Lauga2009}. Moreover, the chaotic bacterial flow inside the drops is indeed isotropic and should average to zero, both spatially and temporally. Therefore, the transfer of motion to the drops is not intuitive. In fact, drops sit over a substrate separated to it by a thin lubrication film, and this spatial symmetry breaking is necessary for the movement of the drops. We show that the drop movement and its direction is determined by the bacteria that move near the substrate, causing the drop to roll over the substrate. The turbulent-like motion of the bacterial bath constantly changes the direction and speed of the bacteria that swim near the bottom of the drop. This explains both the persistent movement of the droplets at short times and their random motion at long times.

In the next section, we present our experimental setup and protocols. Results are presented in Sec.~\ref{sec:Results} and analyzed in Sec.~\ref{sec:Discussion}. Finally, Sec.~\ref{sec:Conclusions} summarizes our main findings.


\section{Experimental setup}

\begin{figure*}[h]
\includegraphics[width=.45\textwidth]{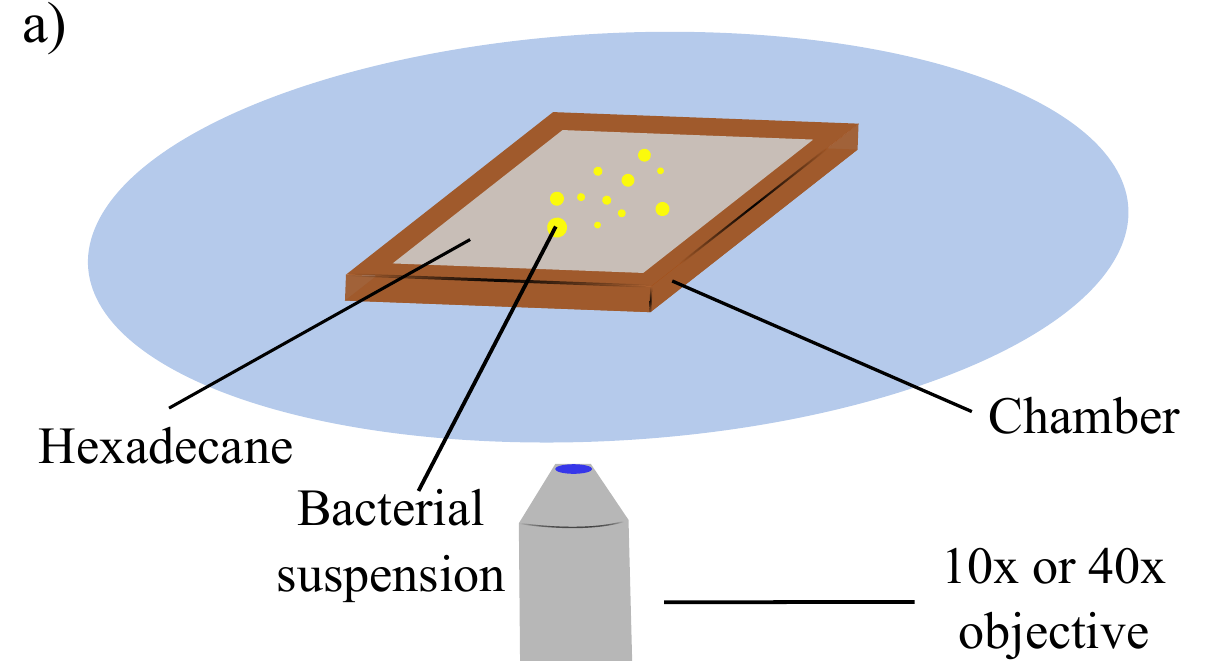}\hfill
\includegraphics[width=.4\textwidth]{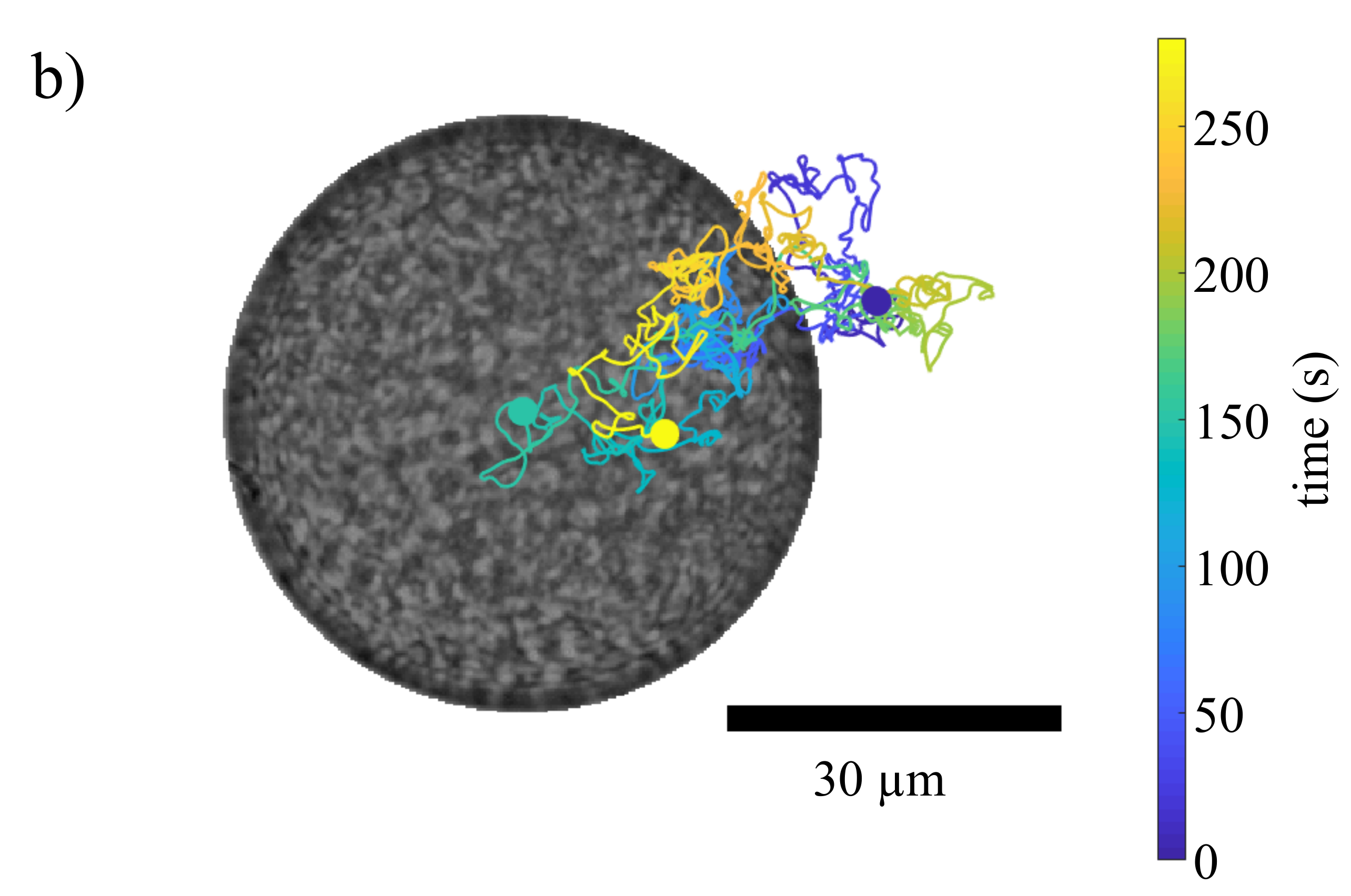}\\
\includegraphics[width=.7\textwidth]{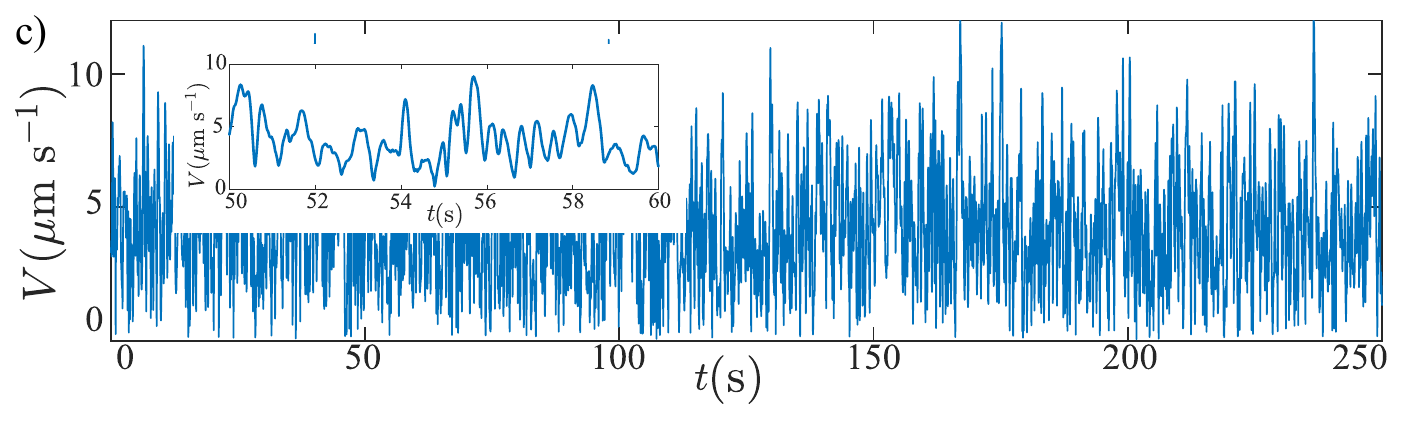}\hfill
\includegraphics[width=.29\textwidth]{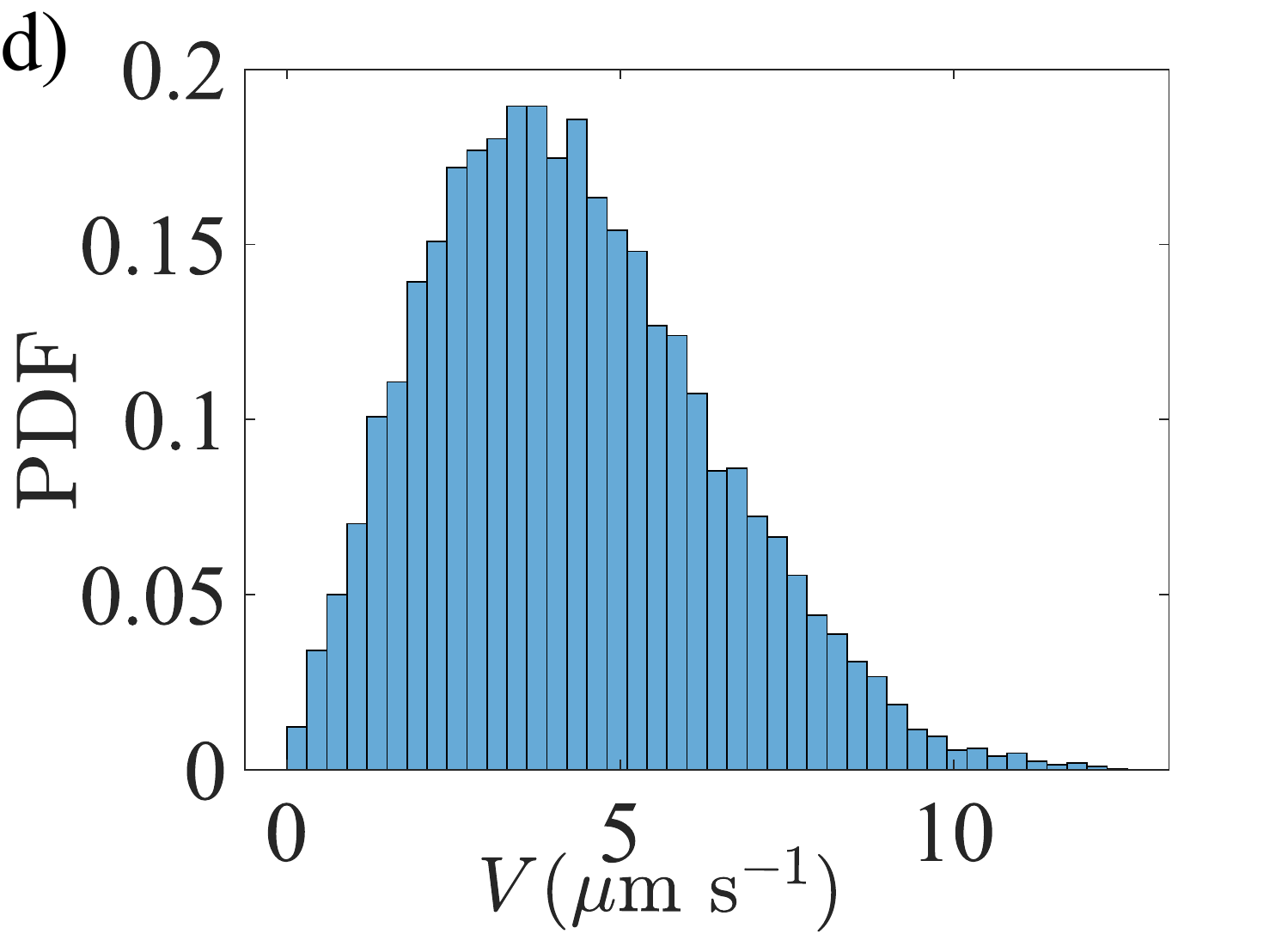}
\caption{a) Schematics of the experimental setup. b) Trajectory of a droplet ($R = \SI{27}{\micro\meter}$ and $n= 1.68 \times 10^{10}~\si{bact/\milli\liter}$) observed in bright field for about 4 minutes. The trajectory is represented by the colored line, with the color scale representing the time from the start (blue dot) to the end (yellow dot). The image of the droplet corresponds to its position \SI{150}{\second} after the initial time (green dot). The high bacterial concentration prevents to distinguish individual \textit{E. coli} inside the drop. c) Instantaneous velocity of the droplet in b). The inset shows a small temporal windows to appreciate the temporal scale of variations. d) Velocity distribution function for the same droplet.}
\label{fig:Setup-DropTrajectory}
\end{figure*}

We use a run and tumble strain of \textit{E. coli} (W3110) which is also genetically modified to express the green fluorescence protein (GFPmut2)\cite{Keymer2008}. Bacteria are collected from a \SI{-20}{\celsius} stock and grown overnight in rich Luria-Bertani medium (LB) at \SI{28}{\celsius} with an agitation of 180~rpm. From this sample, 1 mL of overnight is diluted in 9 mL of LB. The suspension is again incubated at \SI{28}{\celsius} with agitation of 180~rpm and harvested in the exponential phase of the growth curve when it reaches an optical density at 600~nm (OD) of $0.7 \pm 0.1$. The suspension is centrifuged at 3000~rpm for 15~minutes, and the pellet is then washed and diluted in a phosphate motility buffer, MMA (10 mM NaH$_{2}$PO$_{4}$, 10 mM K$_{2}$HPO$_{4}$, 0.1 EDTA and 20 mM sodium lactate)~\cite{Minamino2003}. The motility buffer MMA is a controlled environment where bacteria can swim but do not divide \cite{Altshuler2013}. The concentration of the final bacterial suspension, $n$, ranges from $5.14\times10^{8}$ to $2.25 \times 10^{10}$~bact/mL, depending on the amount of MMA added after centrifugation.

A volume of \SI{10}{\micro\liter} of the bacterial suspension is added to \SI{1}{\milli\liter} of Hexadecane containing Span80 (2$\%$ in weight) as a surfactant. The mixture is manually agitated, resulting in an emulsion of aqueous droplets containing the bacterial suspension in oil. The radii of the droplets, $R$, are widely dispersed, between approximately \SI{10}{\micro\meter} and \SI{200}{\micro\meter}.

The observation setup is a square chamber of inner side $L = \SI{1}{\centi\meter}$ and height $h = \SI{400}{\micro\meter}$. The chamber walls are fabricated in SU-8 photoresist (Gersteltec GM 1075) by optical lithography (Heidelberg Instruments MLA 100) on a \SI{50.8}{\milli\meter} diameter and \SI{500}{\micro\meter} thick circular glass wafer (University Wafer Borofloat 33). After chamber fabrication, the bottom of the chamber is hydrophobically treated using a siliconizing reagent (Sigmacote, Sigma-Aldrich).The emulsion is poured inside the chamber (Fig.~\ref{fig:Setup-DropTrajectory}a) and then the chamber is sealed by a coverslip to minimize evaporation and flows in sample. As the aqueous bacterial suspension is denser than the ambient hexadecane, drops sediment to the bottom of the chamber, but remain separated from the substrate by a thin hexadecane lubrication film, whose thickness we estimate at $\epsilon \approx \SI{20}{\nano\meter}$~\cite{Bretherton1961}. The chamber is placed on an inverted microscope (Nikon TS100F) and filmed by a camera (Andor Zyla $2048\times 2048$ pix$^2$) at 50 fps. Low magnification (10X/0.25 NA objective) in bright field is used to record the movement of several drops simultaneously. High magnification (40X/0.60 NA objective) in fluorescence is used to observe the bacterial flows inside the droplet. 

Image analysis is performed with homemade MATLAB scripts to track the droplets and to analyze the drops trajectories. Particle image velocimetry (PIV) is  performed with the Matlab Toolbox PIVlab \cite{Thielicke2014}, in order to obtain the velocity field inside the drops, as in Refs.~\cite{Vincenti2019} and \cite{Beppu2017}.


\section{Results} \label{sec:Results}

Immediately after the emulsion is poured inside the chamber, observation at the microscope reveals a continuous and erratic movement of the droplets (movie S1). Closer inspection of the drops reveals the existence of collective motion of the bacteria inside the drops, generating a complex inner flow. This flow organizes in vortices that appear, move and disappear continuously, indicative of active turbulence~\cite{wensink2012meso} (movie S2). The movement, both of the drops as a whole and of the bacterial suspension inside the drops, can last several minutes up to several hours. Typically, droplets of radius smaller that \SI{20}{\micro\meter} stop moving after a few minutes, and larger drops move for larger times. We hypothesize that consumption of oxygen or nutrients, or the accumulation of bacterially-generated waste, which happens faster in smaller drops, can cause the bacteria to eventually stop swimming and drops no longer displacing. Figure~\ref{fig:Setup-DropTrajectory}b shows the trajectory of a drop of radius $R = \SI{27}{\micro\meter}$ and bacterial concentration $n= 1.68 \times 10^{10}~\si{bact/\milli\liter}$ for approximately 4 minutes (see also movie S3). Note that the maximum displacement of the drop can be larger than $R$. From the tracked trajectories, it is possible to determine the instantaneous velocity averaging over 5 temporal steps. Figures ~\ref{fig:Setup-DropTrajectory}c-d display the temporal evolution of the speed as well as the measured speed distribution function for the same droplet. The speed is highly fluctuating with a distribution that resembles a Gaussian. Its temporal correlation, which is quantified below, is of the order of a fraction of a second, which is larger than the acquisition time interval \SI{1/50}{\second}, allowing us to extract the instantaneous velocities.

\subsection{Persistent random walk}

\begin{figure}[h]
\centering
\includegraphics[width=.75\linewidth]{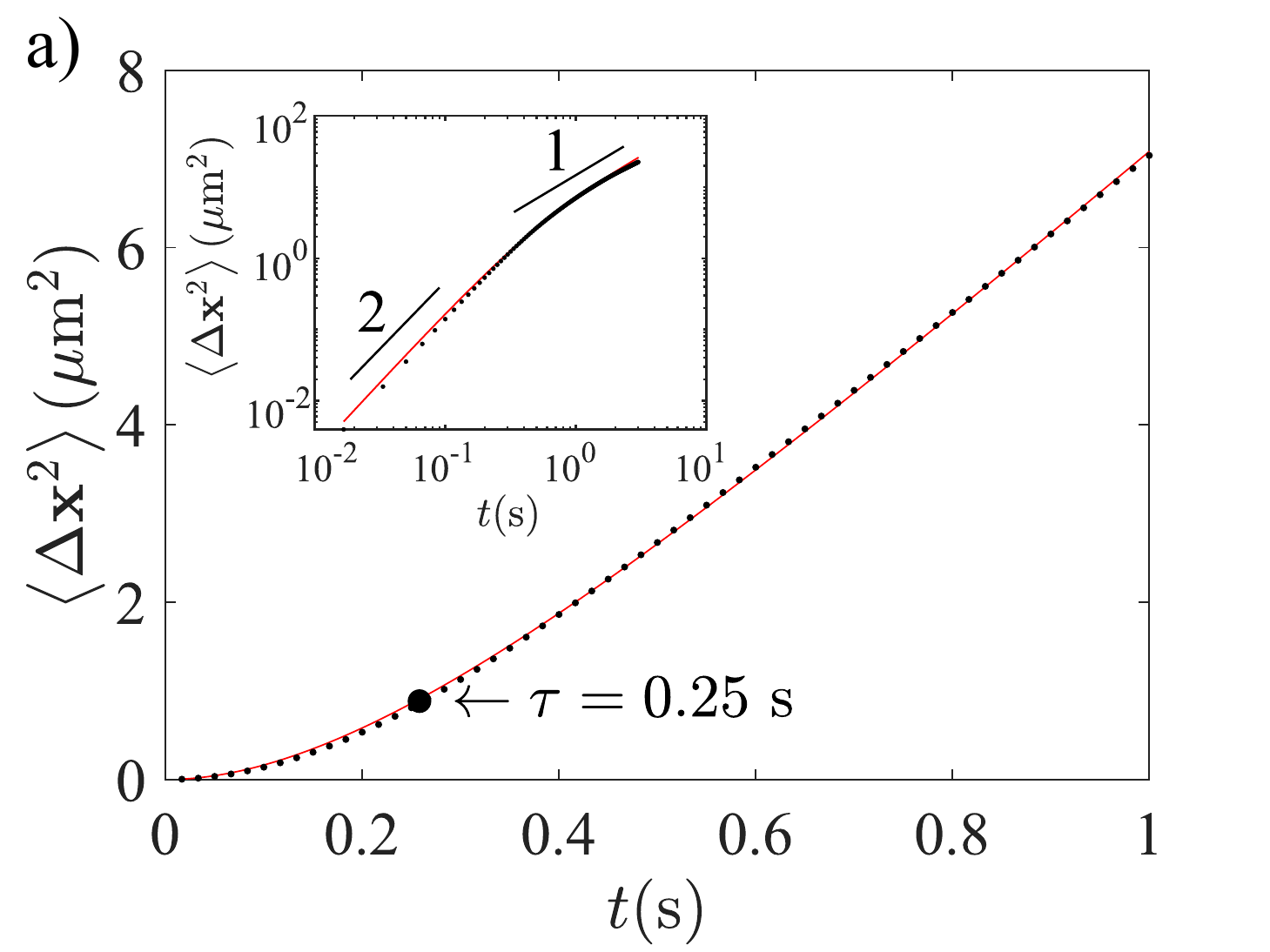}
\includegraphics[width=.75\linewidth]{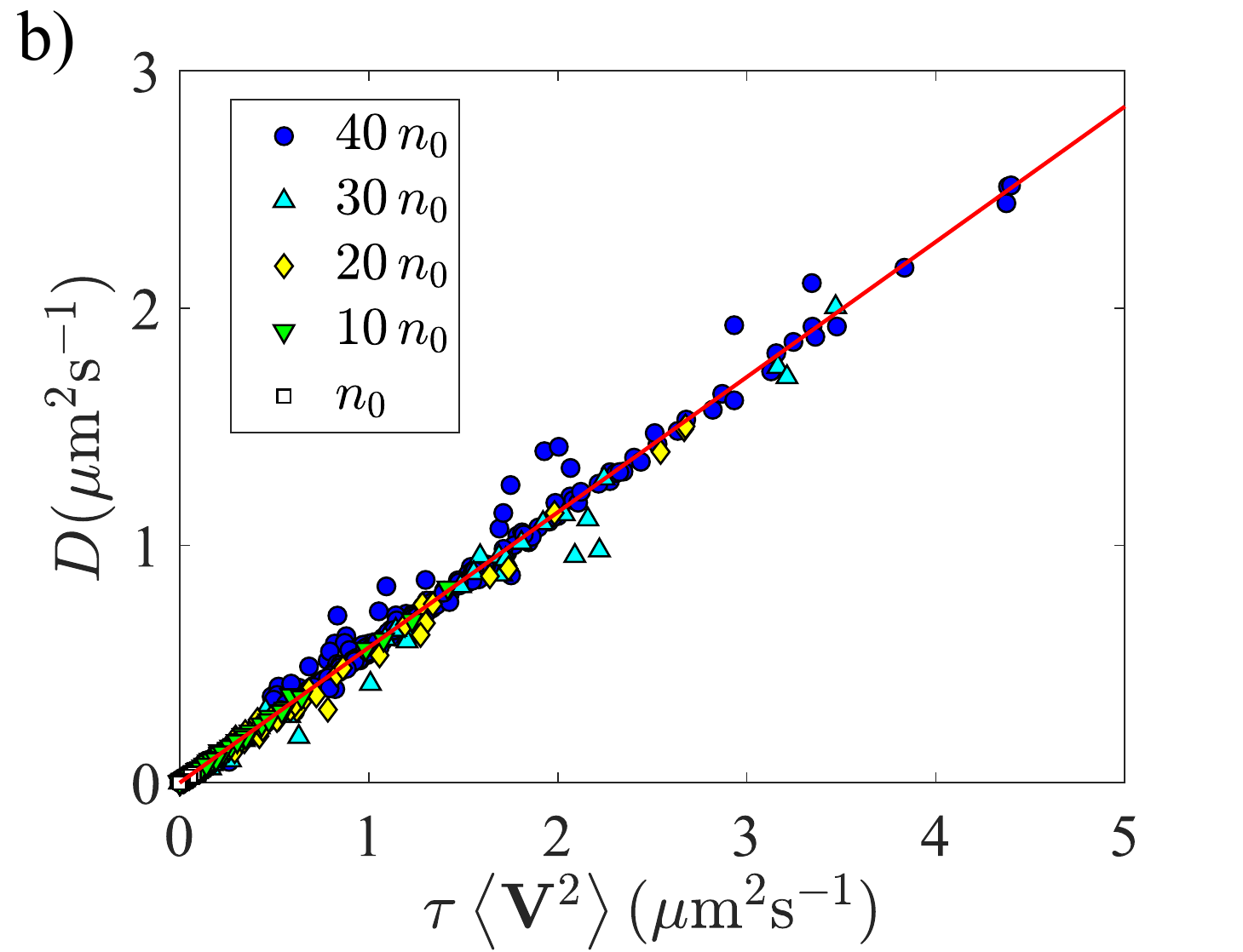}
\caption{a) Representative example of MSD (same droplet as in Fig.~\ref{fig:Setup-DropTrajectory}b). Little black dots are experimental data and the red line is the fit using equation (\ref{eq:ec1}). Persistence time is indicated with the arrow. The inset shows the same data in log-log scale. The two lines are power laws with exponents 2 and 1. b) Measured diffusion coefficient $D$ as a function of $\tau \left\langle \mathbf{V}^{2} \right\rangle$ for all measured drops. The data for different bacterial concentration collapse in a straight line with slope $\alpha = 0.57$ (red line). The reference concentration is $n_0=5.14 \times 10^8~\si{bact/\milli\liter}$.}
\label{fig:MSD-D}
\end{figure}

A total of 991 drops, with radii ranging from $R \approx \SI{20}{\micro\meter}$ to $R \approx \SI{120}{\micro\meter}$  are tracked for 1 minute. Smaller drops are not tracked because they lose their movement too rapidly, and larger drops are discarded to avoid interactions with the upper coverslip. From each droplet trajectory, the mean square displacement (MSD), $\langle\Delta \mathbf{x}^{2} \rangle$, is calculated (Fig.~\ref{fig:MSD-D}a). MSD curves indicate that drops perform a persistent random walk, with a ballistic motion for small times and a diffusive motion at larger times. The shape of the MSD is very well fitted with \cite{Martens2012}
\begin{equation}
\left\langle\Delta \mathbf{x}^{2} \right\rangle = 4D \tau \left(\frac{t}{\tau} - 1 + e^{-t/\tau} \right),
\label{eq:ec1}
\end{equation}
\noindent
where $D$ is the long-time diffusion coefficient of the droplet and $\tau$ the persistence time of the ballistic motion. From eq.~\eqref{eq:ec1}, $D$ and $\tau$ for each droplet are extracted. The instantaneous speeds are averaged over the whole trajectory duration to obtain a mean drop speed, $V$, for each droplet.

The observations present a wide variability; droplets with the same radius and bacterial concentration can present very different values of $V$, $\tau$, and $D$. Despite this variability of the data, individual values of $D$, $\left\langle \mathbf{V}^{2} \right\rangle$ and $\tau$ for each droplet collapse in the line $D = \alpha \tau \left\langle  \mathbf{V}^{2} \right\rangle$ (Fig.~\ref{fig:MSD-D}b).
For a persistent random walk with a velocity autocorrelation function $\left\langle \mathbf{V}(t) \cdot \mathbf{V}(0) \right\rangle = \left\langle  \mathbf{V}^{2} \right\rangle \exp(-t/\tau)$, it is obtained that $\alpha=1/2$ (see Ref.~\cite{Martens2012}). The best fit gives $\alpha=0.57\pm 0.01$. The difference with the persistent random walk prediction suggests that the velocity autocorrelation function is not a single exponential, but the experimental precision does not allow to discriminate among different models.

\subsection{Dependence on bacterial concentration}

\begin{figure*}[t!]
\includegraphics[width=0.33\linewidth]{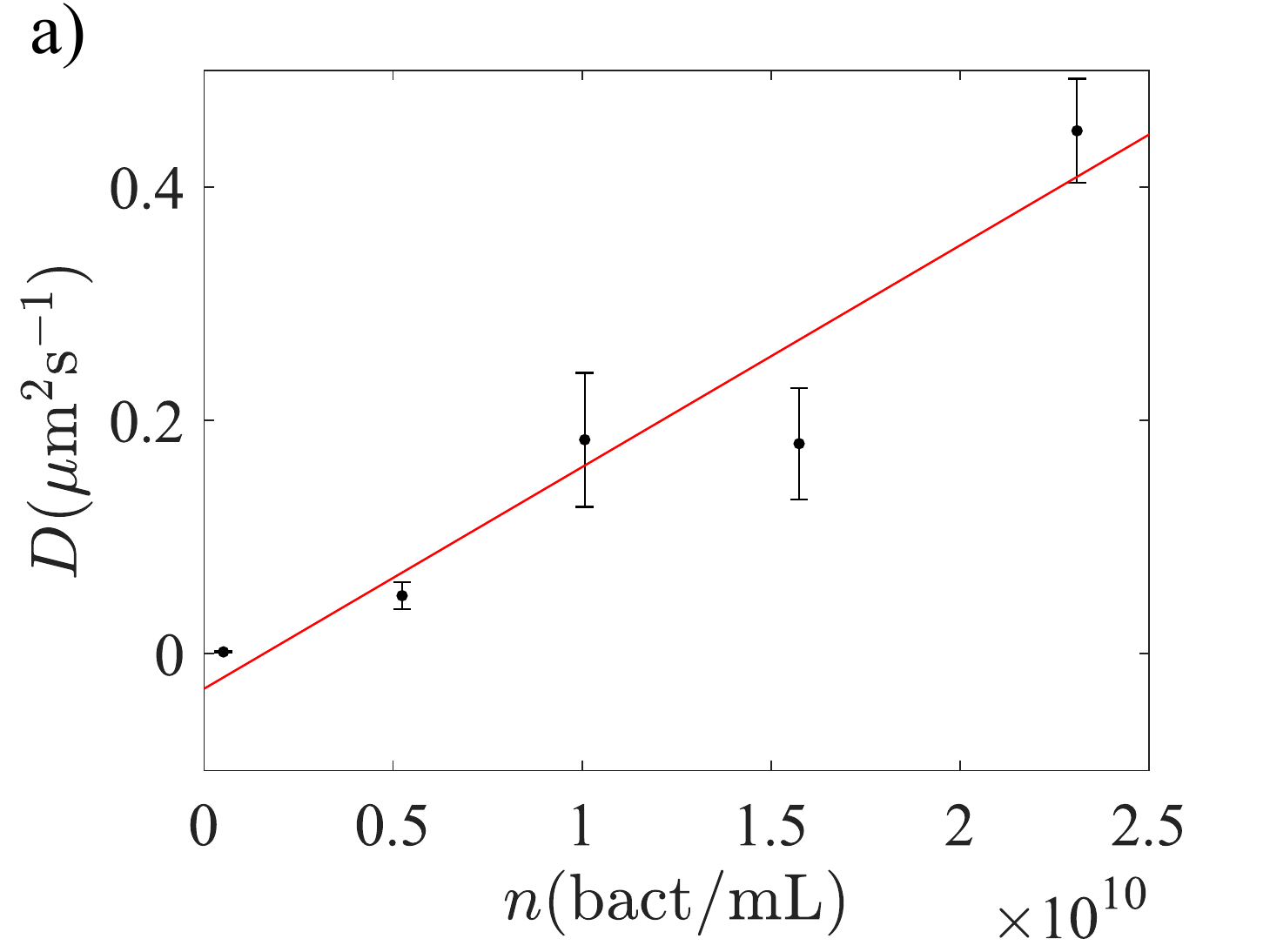}\hfill
\includegraphics[width=0.33\linewidth]{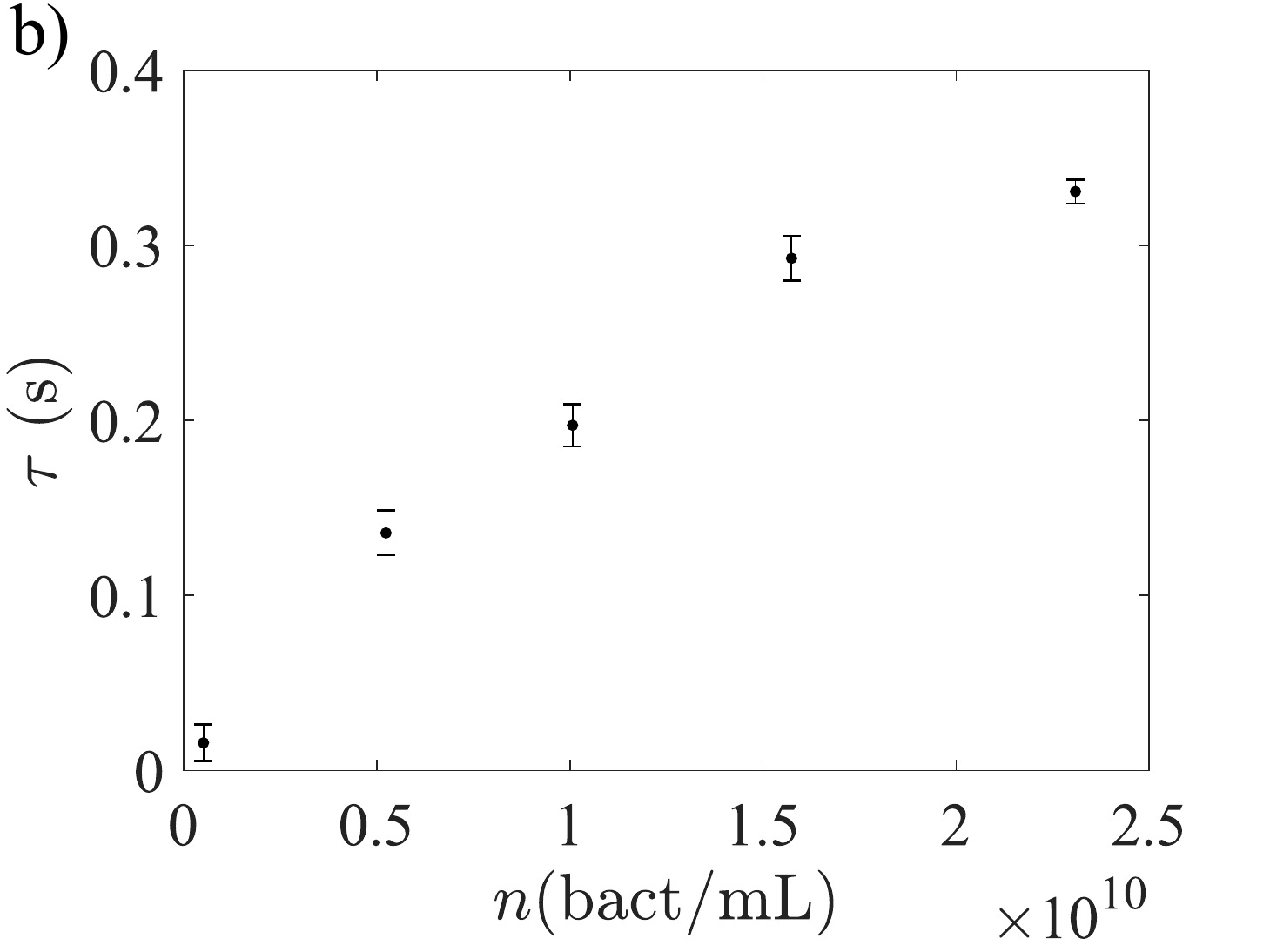}\hfill
\includegraphics[width=0.33\linewidth]{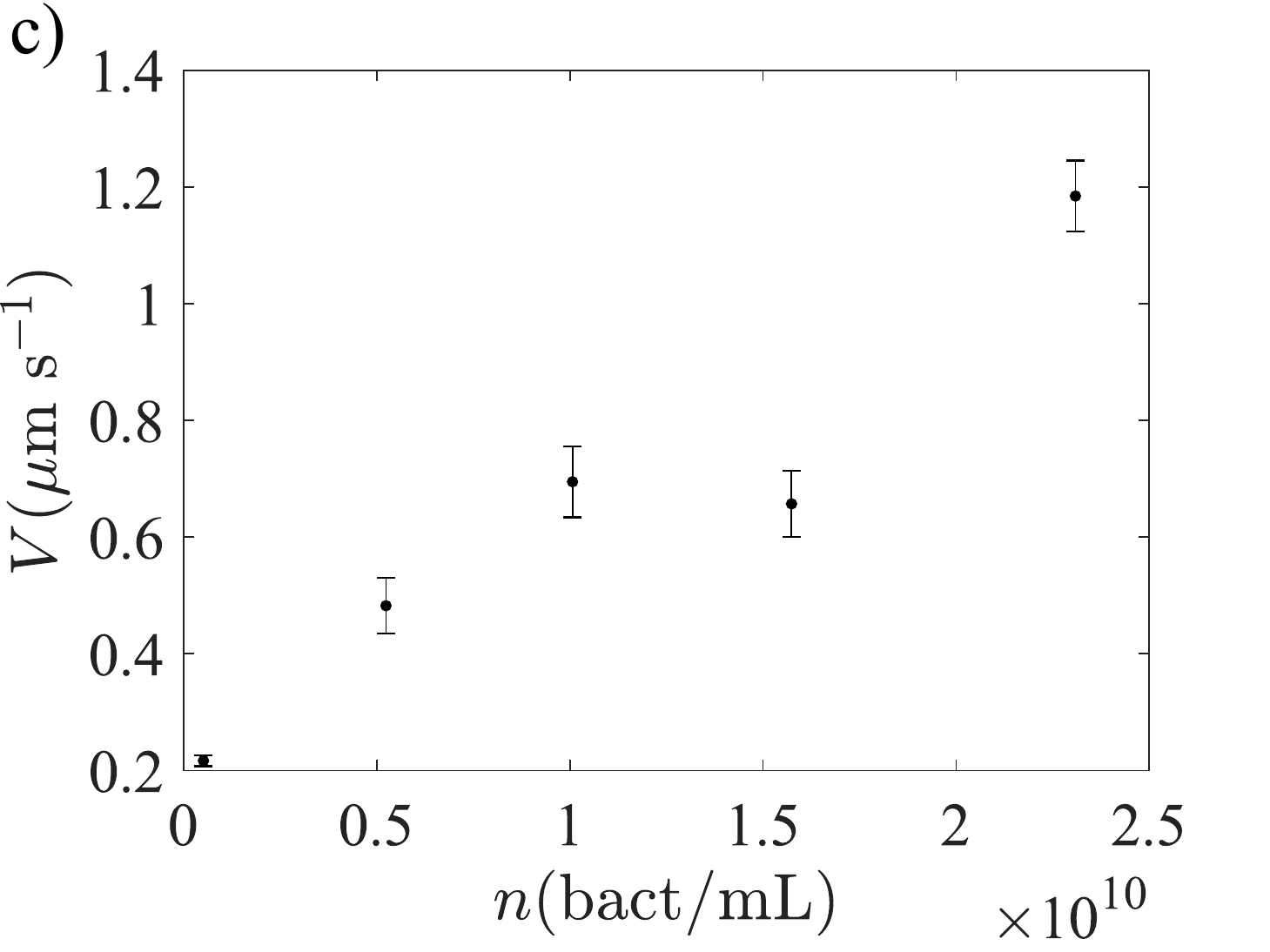}\\
\includegraphics[width=0.33\linewidth]{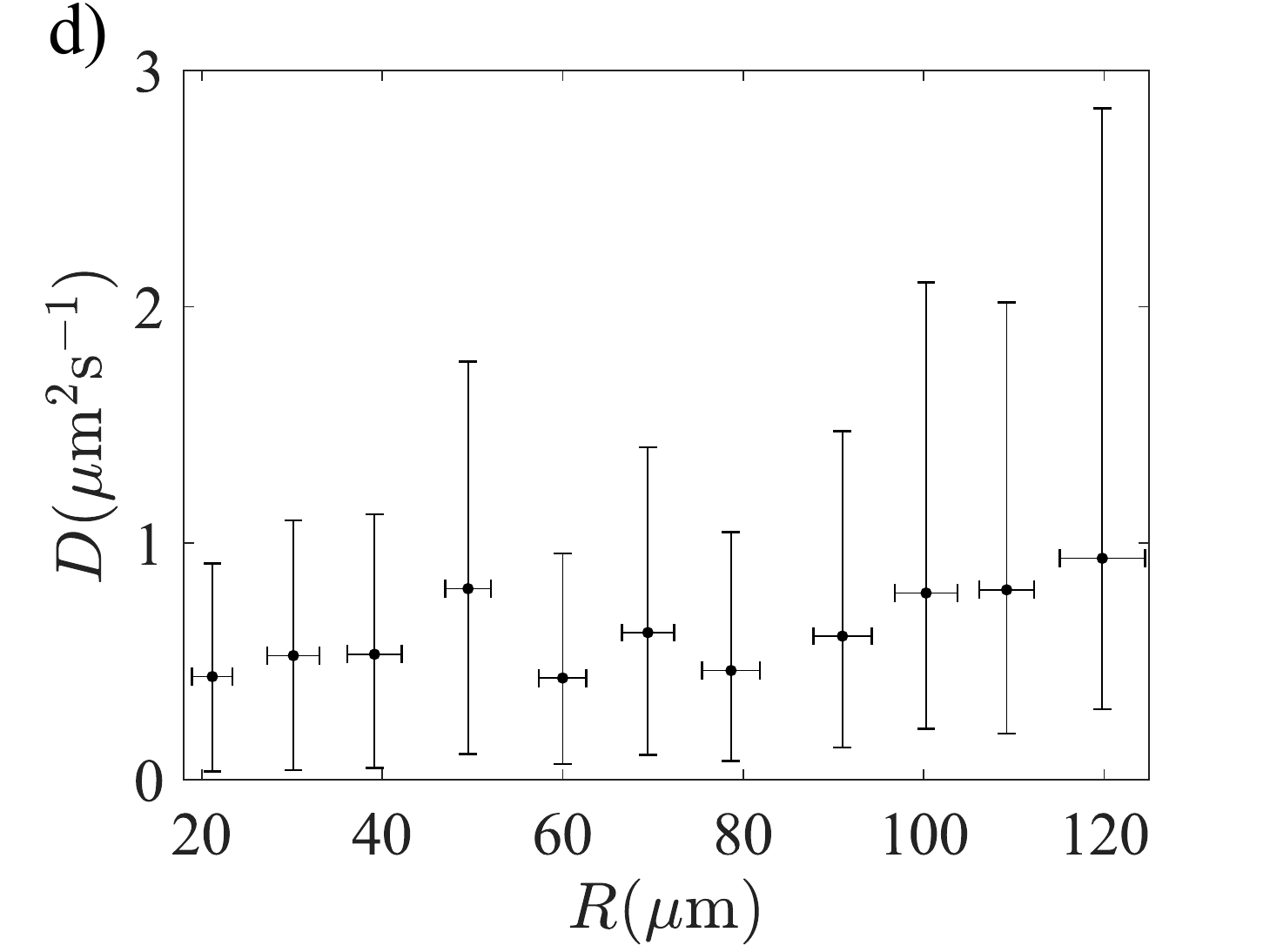}\hfill
\includegraphics[width=0.33\linewidth]{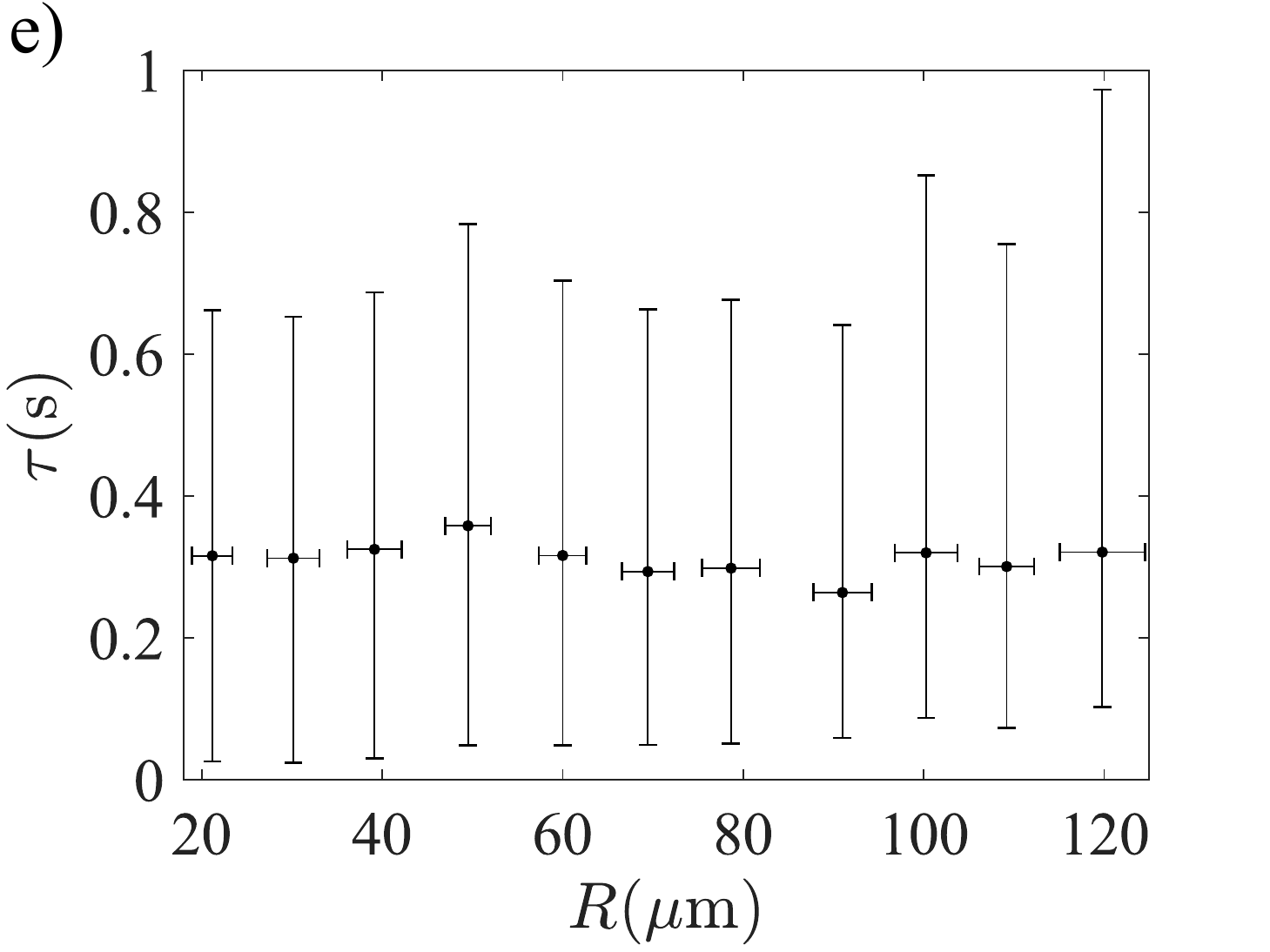}\hfill
\includegraphics[width=0.33\linewidth]{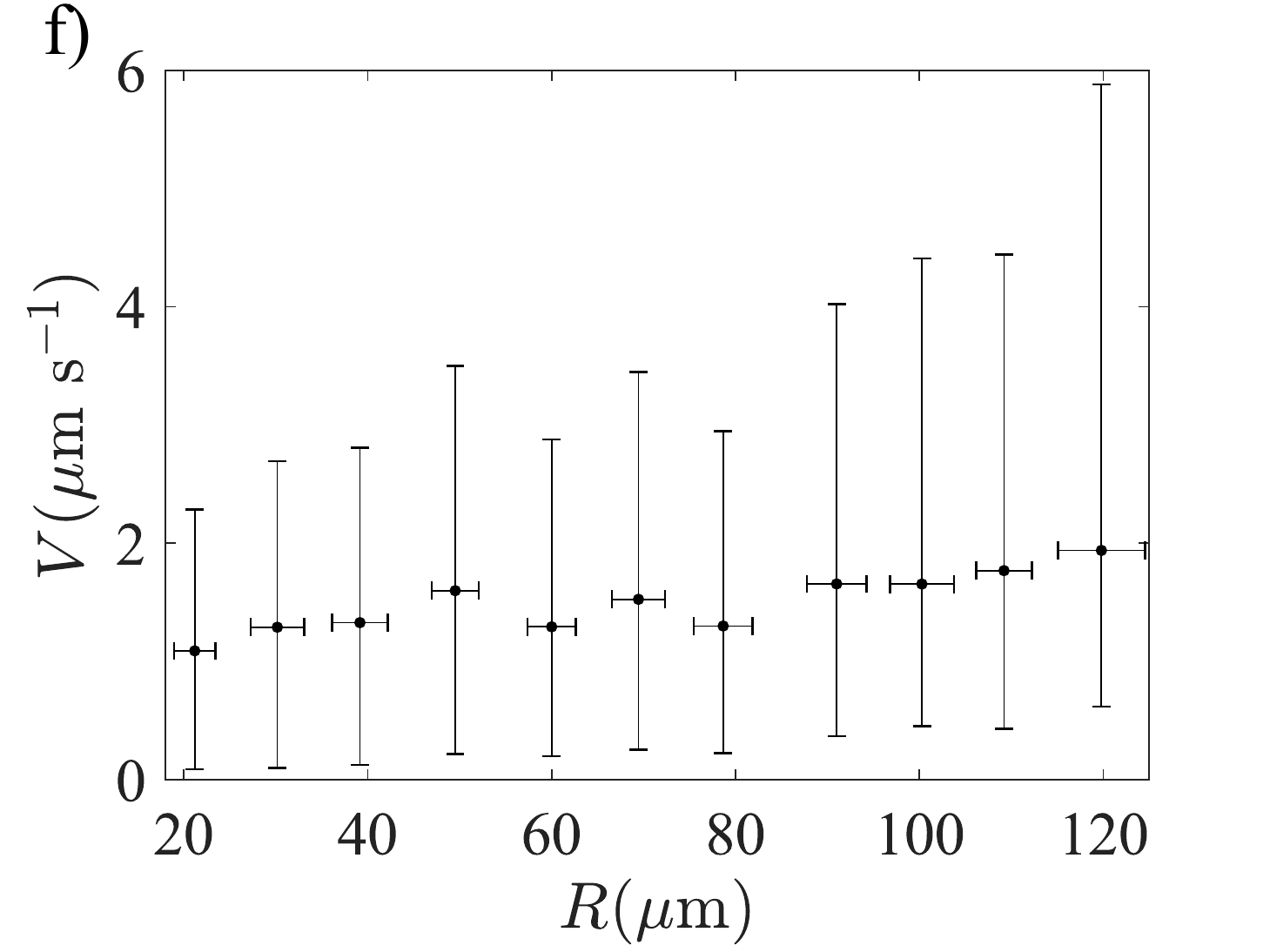}
\caption{ a) Diffusion coefficient, b) persistence time, and c) mean speed as a function of bacterial concentration, averaged for droplets with radii in the range \SI{20}{\micro\meter} to \SI{30}{\micro\meter}.  Symbols represent experimental data and the red line in a) corresponds to a linear fit. d) Diffusion coefficient, e) persistence time, and f) mean speed as a function of droplet radius for fixed bacterial concentration $n = (2.25 \pm 0.14) \times 10^{10}$~bact/mL. The drops are grouped within windows of increasing radius in \SI{10}{\micro\meter} increments. For a-c, the error bars indicate the standard error around the mean value, while in d-e, the horizontal error bars represent the standard deviation on drop radius and the vertical error bar is the confidence interval at 63$\%$.}
\label{fig:Fig3}
\end{figure*}

The movement of droplets was studied as a function of the bacterial concentration $n$ in the bacterial suspension. Droplets with radii from \SI{20}{\micro\meter} to \SI{30}{\micro\meter} were selected, obtaining between 100 and 170 drops for each bacterial concentration. The average diffusion coefficient, persistence time and average speed were calculated at each concentration, as shown in Figs.~\ref{fig:Fig3}a-c. Although dispersion remains high, an increasing tendency of $D$ with $n$ can be observed. Data was fitted to a linear curve $D = D_0+D_1 n$, with 
$D_0 = (-0.03\pm0.15)~\si{\micro\meter^2/\second}$ and $D_1 = (19 \pm 11)~\si{\micro\meter^5/\second/bact}$. The persistence times and the average speed also increase with the bacterial concentration.

\subsection{Dependence on the drop radius}

To analyze the dependence with the droplet radius, we consider the case of maximal bacterial concentration, $n = (2.25 \pm 0.14) \times 10^{10}$~bact/mL, and the drops are grouped within windows of increasing radius in \SI{10}{\micro\meter} increments. The results for $D$, $\tau$, and $V$ as a function of $R$ are presented in Figs.~\ref{fig:Fig3}d-f. For each radius window, a large variability exists for the diffusion coefficient, the persistence time and the average drop speed. The average diffusion coefficient and persistence time remain approximately constant at $D \sim \SI{0.5}{\micro\meter^2/\second}$ and $\tau \sim \SI{0.3}{\second}$. The average speed increases slightly with the drop radius and its average value is  $V \sim \SI{1.5}{\micro\meter/\second}$.

\subsection{Internal flows}

\begin{figure*}[t!]
\includegraphics[width=0.32\linewidth]{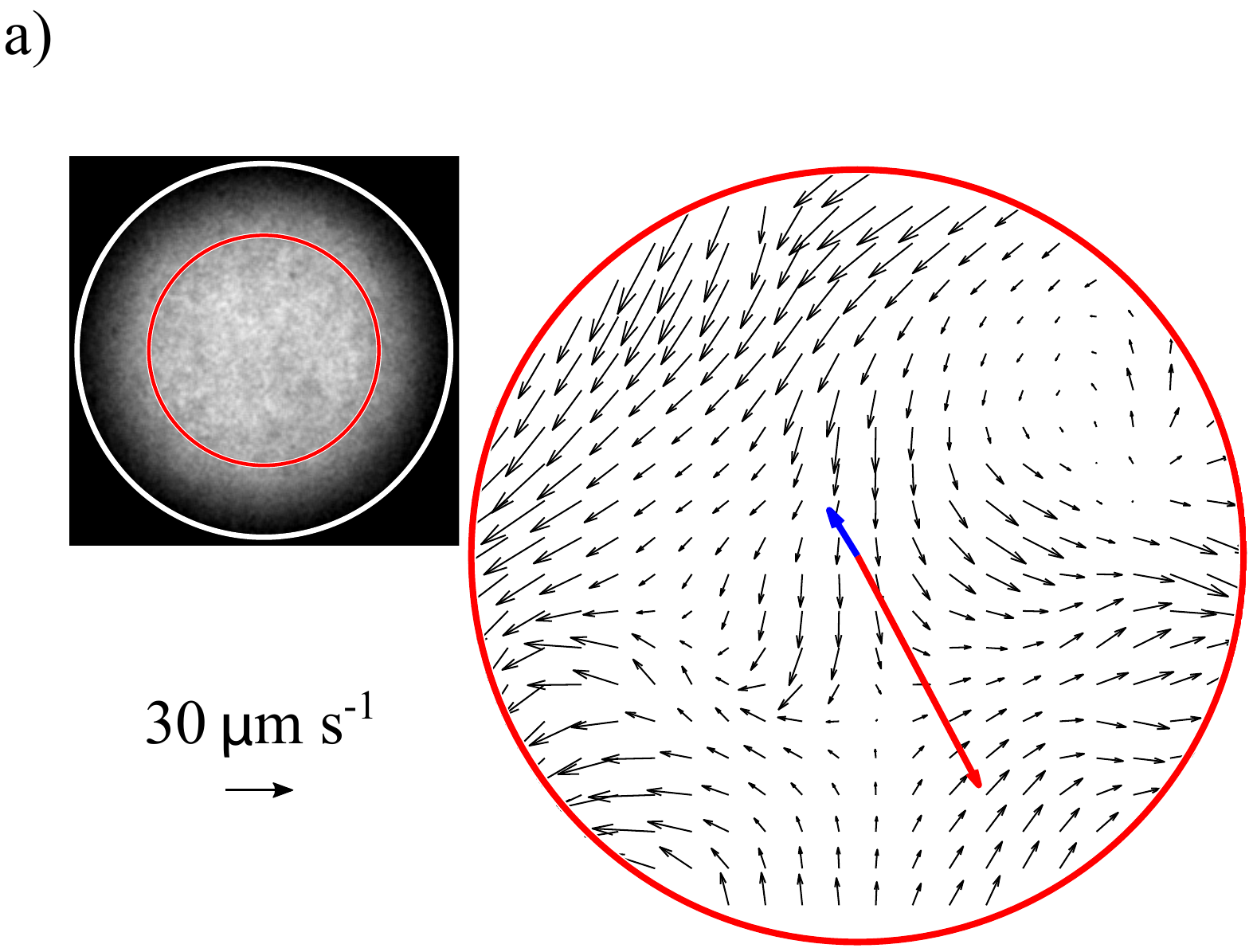}\hfill
\includegraphics[width=0.32\linewidth]{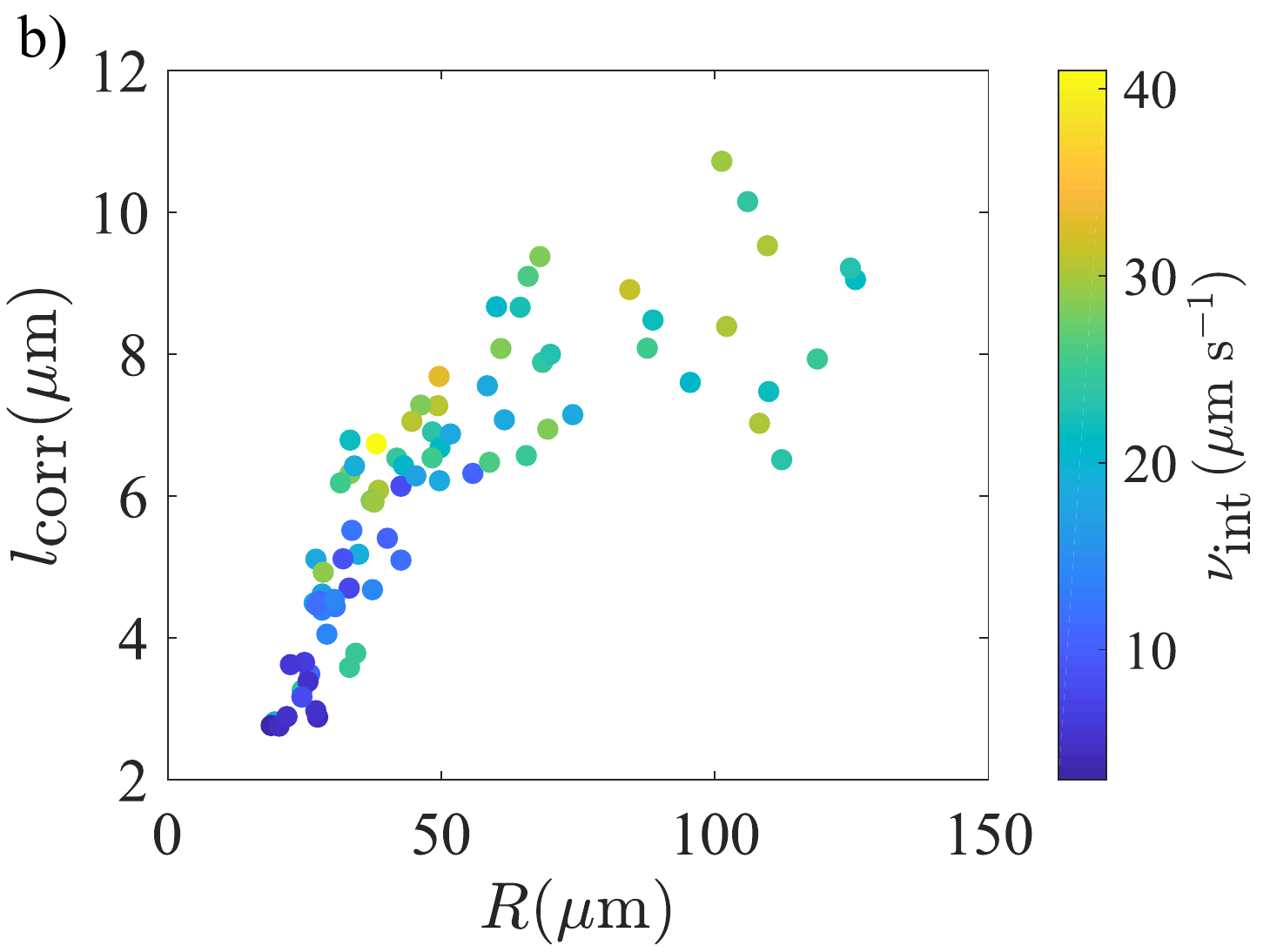}\hfill
\includegraphics[width=0.32\linewidth]{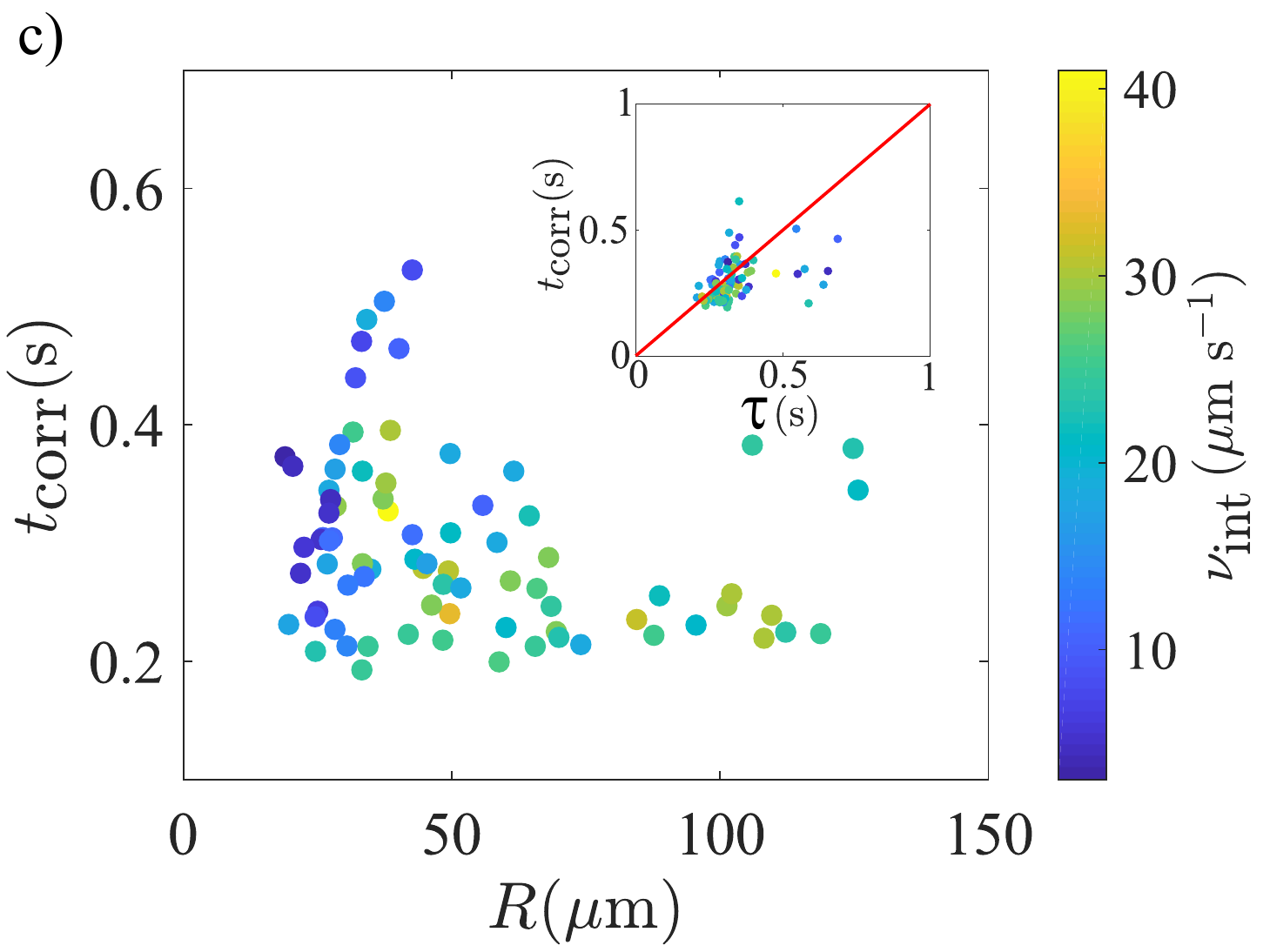}\hfill

\caption{a) Velocity field in the bottom part of a drop, obtained by PIV (black arrows). The red arrow is the weighted average internal velocity $\mathbf{v}_{b}$, while the blue arrow is the instantaneous velocity of the droplet $\mathbf{V}$. For clarity, the red and and blue arrows are enlarged by a factor 10 with respect to the scale. The inset shows a fluorescence image of the bottom part of a drop. The outer white circle marks the boundary of the droplet, of radius $R= \SI{28}{\micro\meter}$. The PIV analysis was performed within the inner red circle of radius $R_{\rm obs}= \SI{18}{\micro\meter}$. b) Correlation length $l_{\rm corr}$ and c) correlation time $t_{\rm corr}$ for the velocity field inside the droplets as a function of drop radius $R$ and mean speed of the internal flow, $v_{\rm int}$ (color scale). Inset: Correlation time of the internal flow, $t_{\rm corr}$, compared with the persistence time of the drop motion, $\tau$. The red straight line has slope one and is a guide to the eye.}
\label{fig:Fig4}
\end{figure*}

Internal flows in the bottom of the drops were recorded in fluorescence near the contact surface with the substrate. Fluctuating vortical flows driven by the bacterial activity are observed (movie S3). Velocity fields $\mathbf{v}(x,y,t)$ are obtained with PIV in a small region of the bottom of the drops, corresponding to the brightest portion of the image. For calculation purposes, we chose an observation region of radius $R_{\rm obs} = \sqrt{7}R/4 $ that corresponds to a height $h_{\rm obs} = R/4$ measured from the bottom (Fig.~\ref{fig:Fig4}a). 

Several drops are measured and, from each velocity field, the mean speed of the internal flow $v_{\rm int}\equiv\langle |\mathbf{v}(x,y,t)|\rangle$ is computed, where the average is over time and space. Also, the average over time of the correlation length  $l_{\rm corr}$ for the velocity field is calculated as described in Ref.~\cite{Sokolov2012}. The dependence of $l_{\rm corr}$ on drop radius $R$ and mean internal speed $v_{\rm int}$ is presented in Fig.~{\ref{fig:Fig4}b. For drop radius between \SI{20}{\micro\meter} and \SI{50}{\micro\meter}, $l_{\rm corr}$ increases rapidly from $\approx \SI{3}{\micro\meter}$ to $\approx \SI{7}{\micro\meter}$, with relatively little dispersion. For larger drop radius, $l_{\rm corr}$ seems to plateau at a value of $\approx \SI{8}{\micro\meter}$, but with larger dispersion. The dependence of $l_{\rm corr}$ on $v_{\rm int}$ (color scale of Fig.~\ref{fig:Fig4}b) presents a larger dispersion, but in general  $l_{\rm corr}$ increases with $v_{\rm int}$.

Finally, the correlation time of the velocity field, $t_{\rm corr}$, was obtained as follows. For each position in the observation region, the temporal self-correlation function of the velocity is obtained. From it, we compute the first moment, and the average of these give $t_{\rm corr}$. Figure \ref{fig:Fig4}c shows that the correlation time is almost insensitive to the droplet radius, with a mean value of $t_{\rm corr} = (0.30 \pm 0.08) \si{\second}$.


\section{Discussion} \label{sec:Discussion}
\subsection{Bacterial activity enhances droplet diffusion} 

Thermal diffusion of a sphere of radius $R$ in a medium of viscosity $\eta$ at temperature $T$ is given by $D_{\rm th} =k_{\rm B}T/(6\pi \eta R)$. In the case of a liquid drop of viscosity $\eta'$, a  factor $C(\eta'/\eta)$ must be included, accounting for the slip condition at the drop surface, $D_{\rm th} = k_{\rm B}T/[C(\eta'/\eta) \eta R]$ \cite{Bond1927}. Using $\eta' = \SI{1}{\milli\pascal\second}$ for water and $\eta = \SI{3.47}{\milli\pascal\second}$ for hexadecane, gives $C=13.9$. With $T = \SI{20}{\celsius}$ and $R = \SI{20}{\micro\meter}$ (the smallest radius considered here), we obtain $D_{\rm th} \approx \SI{0.4e-2}{\micro\meter^2/\second}$. This value is two orders of magnitude smaller than the average diffusion coefficient $D \approx \SI{0.3}{\micro\meter^2/\second}$ that we measure for drops at the higher bacterial concentration used here. For a droplet of mass $M$, its stopping time in hexadecane is $\tau_{\rm stop}=M/[C(\eta'/\eta) \eta R]$. The same parameters used to estimate $D_{\rm th}$ give $\tau_{\rm stop}=\SI{3.3e-5}{\second}$, which is much smaller than the persistence time $\tau$ of the ballistic motion of the drops.

Our results do not evidence any significant dependence of the diffusion coefficient with the drop radius (Fig.~\ref{fig:Fig3}d), as opposed to the $1/R$ scaling of $D_{\rm th}$. This is also true for the mean persistence time $\tau$ (Fig.~\ref{fig:Fig3}e). The mean droplet speed $V$ (Fig.~\ref{fig:Fig3}f), exhibits a slight increase with the drop radius that could be due to the loss of bacterial activity, which occurs faster for smaller droplets.

Finally, diffusivity of the droplets increases linearly with the concentration of the bacterial suspension (Fig.~\ref{fig:Fig3}a). The intercept found with the linear fit is consistent with the thermal diffusivity $D_{\rm th}$ expected for drops in the size range studied here.

These observations indicate that bacterial activity is the motor of the movement that we observe in the drops. In the following we present a model to explain the drop movement based on the internal flows driven by the bacteria.

\subsection{Driving mechanism}

In dense suspensions, swimming bacteria usually organize in collective motions. Inside the spherical drops that we study, these collective motions translate in vortices that appear, move and disappear continuously, with a characteristic size $l_{\rm corr}$ and life time $t_{\rm corr}$. The value of $l_{\rm corr}$ shows an increasing tendency with the drop radius, probably related to confining effects (Fig.~\ref{fig:Fig4}b). Comparison of the characteristic duration of these collective motions with the persistence time of the ballistic motion of the droplet (Fig.~\ref{fig:Fig4}c, inset) reveals that both are restricted to the same range, around \SI{0.3}{\second}.

\begin{figure*}[h]
\raisebox{0.1\height}{\includegraphics[width=0.32\linewidth]{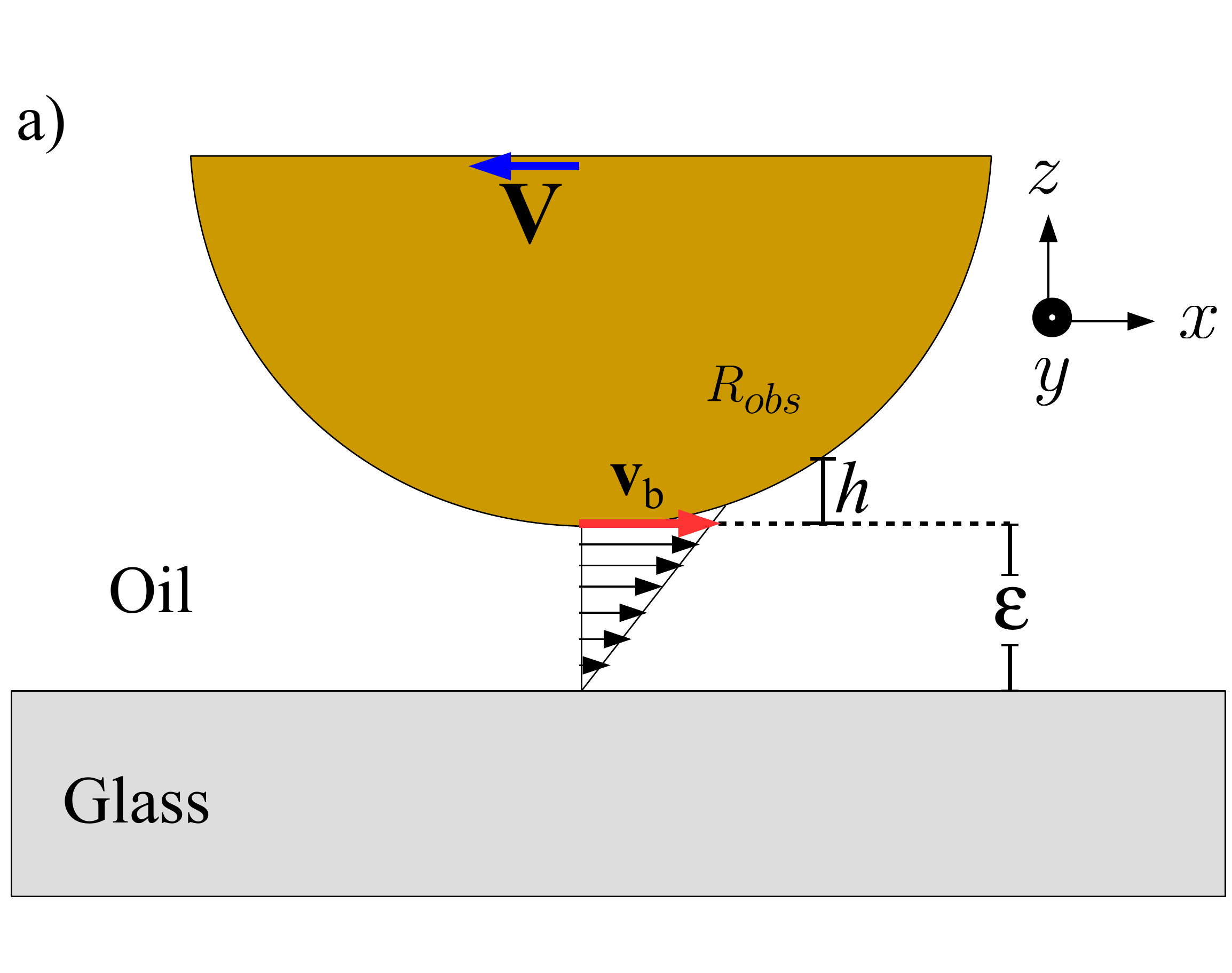}}\hfill
\includegraphics[width=0.32\linewidth]{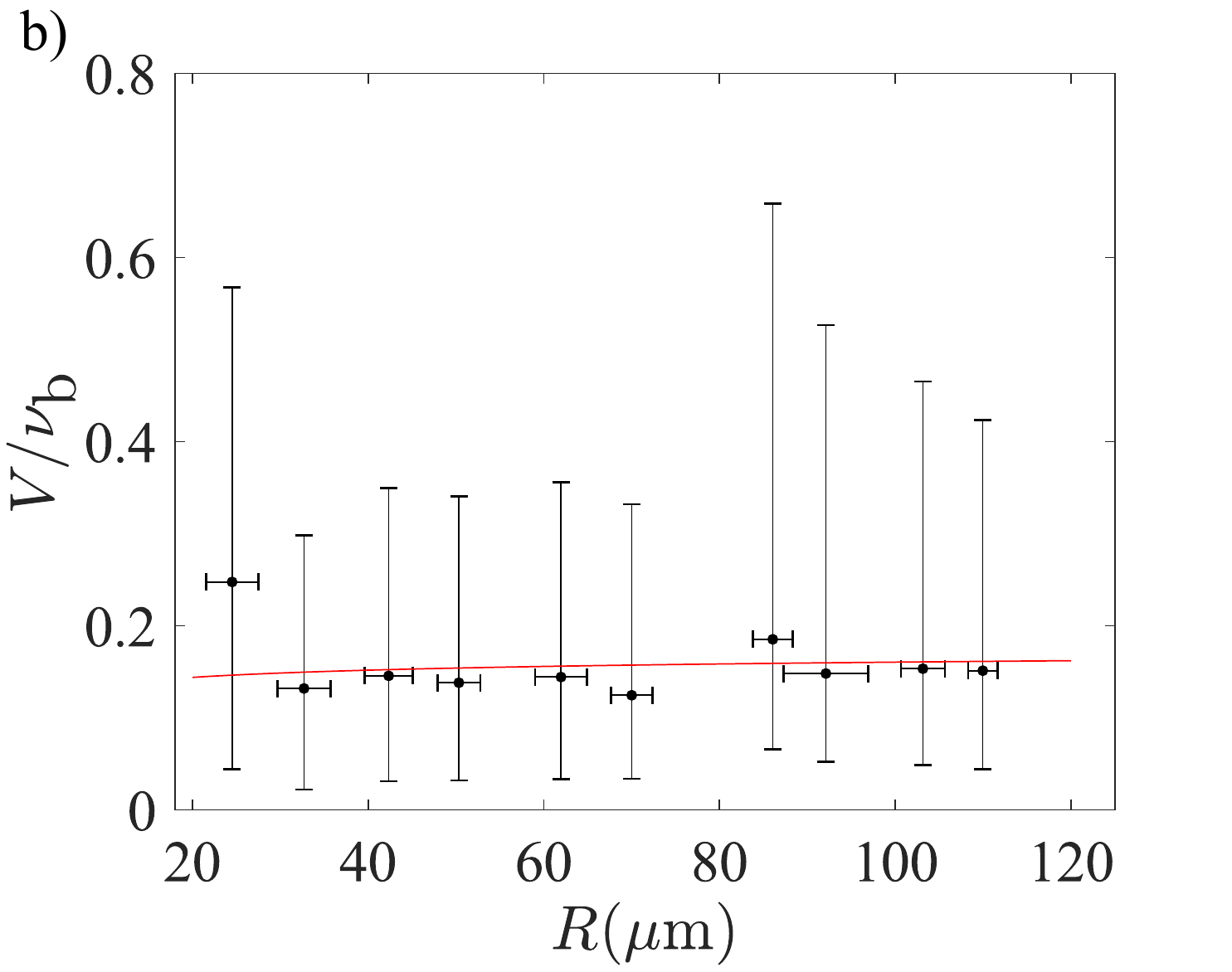} \hfill
\includegraphics[width=0.34\linewidth]{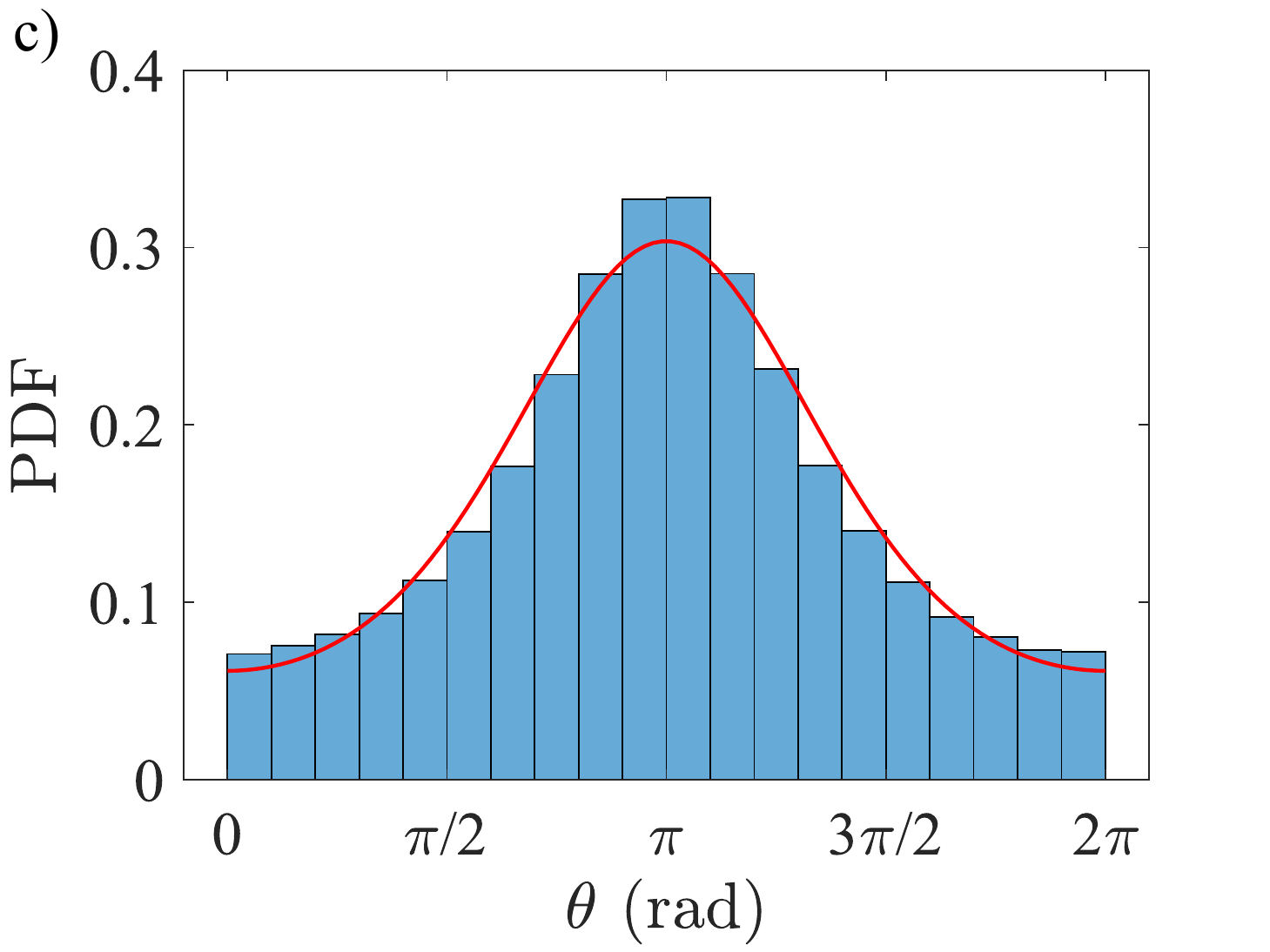} 
\caption{a) Schematics of the rolling with slipping of a drop. $\mathbf v_{\rm b}$ is obtained averaging the velocity field in the observation region delimited by $R_{\rm obs}$ and averaged with the weight function $w$ (eq.~\ref{eq.weigth}), which depends on the relative height to the bottom of the droplet $h$. The thickness of the lubrication film $\epsilon$ is shown out of scale for clarity. b) $V/v_{\rm b}$ as a function of drop radius $R$. Symbols represent experimental data, averaged over drops within windows of increasing radius in \SI{10}{\micro\meter} increments. Horizontal error bars correspond to standard deviation on drop radius while vertical error bars represent the confident interval at 63\%. Red curve corresponds to eq.~(\ref{eq:RollingModel}). c) Probability density function for the angle between $\mathbf{v}_{\rm b}$ and $\mathbf{V}$. The red line is the fit for the von Mises distribution, eq.~(\ref{eq:vonMises}), showing a peak at $\theta = \pi$.}
\label{fig:Fig5}
\end{figure*} 

Experimental results indicate that the bacterial motion inside the drops is responsible for the movement of the drops. The bacterial currents are highly fluctuating, with  correlation lengths smaller than the droplet radii. Hence, if the droplet were suspended in an infinite fluid medium, the drag forces on different regions of the droplet surface would almost cancel out, resulting  in small instantaneous values.
We hypothesize that the presence of the bottom glass substrate, on the contrary, enhances the drag effect of these fluctuating currents. In effect, only a small region of the droplet is in contact with the substrate, where it is more probable to observe coherent motion. As the friction close to the solid surface becomes larger, the cancelation of the drag forces described above does not take place.
 The mechanism that we propose is a ``rolling with slipping''. When the bacterial motion inside the drops produces a patch of directed flow in the lowest part of the drop, the thin lubrication film existing between the glass substrate and the drop is sheared in the direction of the flow. This shear, in turn, creates a net force on the drop that causes its movement in the opposite direction of the inner flow, as schematically shown in Fig.~\ref{fig:Fig5}a.
By lubrication theory, we can approximate the motion of the bottom part of the droplet as having an angular velocity $\Omega=v_{\rm b}/R$, where $v_{\rm b}$ is the speed of the inner flow at the bottom of the drop. Note that we are not assuming that the droplet rotates as a whole because only  the bottom region of the droplet is relevant for the lubrication theory calculation. Considering that the droplet moves at a velocity $\mathbf V$, the total hydrodynamic force on the droplet is $\mathbf F^\text{hydro} = \mathcal R^{FU} \mathbf V +\mathcal{R}^{F\Omega}\boldsymbol\Omega$, where  $\mathcal R^{FU}$ and $\mathcal R^{F\Omega}$ are resistance tensors, which depend on the lubrication layer thickness $\epsilon$ \cite{dunstan2012two}. At vanishing Reynolds number, the hydrodynamic force cancels, resulting in
\begin{equation}\label{eq:RollingModel}
\mathbf{V} = - \frac{\mathcal R^{F\Omega}_{/\mkern-5mu/}}{R \mathcal R^{FU}_{xy}} \mathbf v_{\rm b} 
= -\left(\frac{2 \ln(\epsilon/R)/15+0.2526}{8 \ln(\epsilon/R)/15- 0.9588}\right) \mathbf v_{\rm b},
\end{equation}
where we used the expressions for the resistance tensors in Ref.~\cite{dunstan2012two}. In eq.~(\ref{eq:RollingModel}), $\mathbf v_{\rm b}$ is the relevant velocity at the bottom of the drop, caused by the bacterial motion and the minus sign indicates that the drop velocity $\mathbf V$ is antiparallel to $\mathbf v_{\rm b}$.

To test this hypothesis, we define a weighted average internal velocity $\mathbf{v}_{\rm b}$, with a weight function $w(x,y)$,
\begin{equation}
w(x,y) = \frac{1}{h(x,y)^{2} + \epsilon^{2}}, \label{eq.weigth}
\end{equation}
where $(x,y)$ represent the horizontal coordinates of a point in the PIV velocity field, $h(x,y)$ is the height from the lowest point of the droplet to the interface at position $(x,y)$ (see Fig.~\ref{fig:Fig5}a). The square in $h$ is used to give more weight to the points near the glass surface and roughly models the decay of the flow produced by force dipoles. $\epsilon$ is a regularizer to avoid divergences and is estimated to be of the order of the  lubrication film thickness. We take $\epsilon = \SI{20}{\nano\meter}$. Then, 
\begin{equation}
\mathbf{v}_{\rm b}(t) =  \frac{\sum_{x,y} \mathbf{v}(x,y,t) w(x,y)}{\sum_{x,y} w(x,y)}.
\end{equation}
This weighted average velocity is compared with the velocity of the droplet $\mathbf{V}$ obtained by the drop tracking. Figure~\ref{fig:Fig5}b shows our experimental results of $V/v_{\rm b}$ as a function of $R$, together with the slippery rolling model, eq.~(\ref{eq:RollingModel}). Despite the variability in the experimental data, the agreement between the experiment and the model is very good, except for the smallest droplets.

In general, $\mathbf{v}_{\rm b}$ and $\mathbf{V}$ go in opposite direction, as shown by the red and blue arrows in Fig.~\ref{fig:Fig4}a, respectively. To quantify this antiparallel behavior, the angle $\theta$ between $\mathbf{v}_{\rm b}$ and $\mathbf{V}$ is determined in the range $\left[0,2 \pi \right]$ along the whole trajectory for all droplets. The probability density function (PDF) of $\theta$ is presented in Fig.~\ref{fig:Fig5}c. A peak near $\theta = \pi$ is clearly evident. The location of the peak and the width of the distribution is determined by fitting a von Mises distribution
\begin{equation}
\label{eq:vonMises}
P(\theta) = \frac{1}{2 \pi I_{0}(k)}e^{k\cos(\theta-\theta_0)},
\end{equation}
where $I_{0}$ is the modified Bessel function of the first kind and $\theta_0 = \pi +\delta$ is the position of the center of the distribution, with $\delta$ the deviation from the perfect antiparallel alignment between $\mathbf v_{\rm b}$ and $\mathbf V$. Finally, $k$ is a measure of the width of the distribution. For $k = 0$ the distribution is uniform and for $k \ne 0$ it is more concentrated in a certain angle. Fitting of the PDF with the von Mises distribution yields $k = 0.80$ and $\delta=-0.0015$ (Fig.~\ref{fig:Fig5}c, red line). Since $\delta \ll \sqrt{k}$ and $P(\pi)/P(0)=\exp(2k)\gg1$,  the vectors are, on average,  effectively antiparallel.


\section{Conclusions} \label{sec:Conclusions}

In this work we have shown that dense suspensions of motile bacteria encapsulated inside emulsion drops are able to transfer movement to the drops. These bacterially propelled drops perform a persistent Brownian motion, which we have systematically studied as a function of the bacterial concentration and drop radius. 
The diffusion coefficient, average speed and persistence time of the droplets present a wide variability. A possible origin of this variability is that, although the prepared suspensions have well controlled bacterial concentrations, the droplet production by agitation could induce concentration inhomogeneities, resulting in droplets of unequal concentration. Further studies are needed to test this hypothesis.

We have shown that bacterial coordinated activity and the presence of a substrate are essential for the propulsion of the droplets. We demonstrate that bacteria drive the droplet by comparing the velocity of the center of mass of the drop with a relevant average velocity of the bacterial suspension in the bottom of the drop, and also comparing the life time of the bacterial collective motions with the persistence time of the droplet motion in the ballistic regime.

In this way we show that the encapsulation of active particles can create another active particle, a motor made of motors~\cite{Sanchez2012}. The power extracted from this motor is quite low; for a drop of typical radius $R \sim \SI{50}{\micro\meter}$ we estimate it at $P_{\rm drop} = C(\eta'/\eta)\eta R V^2 \sim \SI{0.01}{\femto\watt}$, comparable to previous works~\cite{DiLeonardo2010}, but three orders of magnitude below the maximum power available from the bacterial bath for the corresponding drop volume of \SI{5e-7}{\milli\liter} ($\sim \SI{50}{\femto\watt}$). Of course, this leaves ample room for improvement, however, most of the bacterial power will be dissipated as viscous heating, limiting the efficiency of mesoscopic bacterial motors.

Some limitations remain a challenge due to the biological properties of the living particles, for example, the loss of activity over time. Its inhibition through an appropriate regenerating system for the chemical environment should enhance the droplets lifetime, as proven in Refs.~\cite{Sanchez2012, Keber2014}. In comparison with these experiments with microtubules and kinesin molecular motors, {\it E. coli} are simple to culture and handle, representing a widespread experimental model for active matter. In this case, the use of mutant strains with higher resistance to oxygen depletion and/or accumulation of detritus~\cite{Keymer2008} should prove beneficial. Moreover, chemotactic behavior of bacteria represents an interesting possibility to control the drop trajectory by external, manipulable fields. Thus, bacteria could be used to transport components within the droplet, or even to transport themselves and produce, in the right place, other components of interest, like proteins or enzymes to be used in medical treatments or biochemical processes.  


\section*{Acknowledgements}
The authors would like to thank E. Clement for valuable discussions and J. Keymer and J. Noorlag for providing us with the {\it E. coli} strain and assistance in the culture protocols. G.R. thanks CONICYT grant Doctorado Nacional 21150648. This work is supported by the Millenium Nucleus Physics of Active Matter of the Millenium Scientific Initiative of the Ministry of Economy, Development and Tourism (Chile), and Fondecyt grants No. 1180791 and 1170411. Observation chambers were fabricated in the Laboratory of Optical Lithography, built thanks to Fondequip grants EQM140055 and EQM180009.



\balance



\providecommand*{\mcitethebibliography}{\thebibliography}
\csname @ifundefined\endcsname{endmcitethebibliography}
{\let\endmcitethebibliography\endthebibliography}{}

\end{document}